\documentclass[twocolumn,showpacs,preprintnumbers,amsmath,amssymb,superscriptaddress]{revtex4}

\usepackage{graphicx}
\usepackage{dcolumn}
\usepackage{bm}

\begin{document}


\title{Time-reversal-protected single-Dirac-cone topological-insulator
states in Bi$_2$Te$_3$ and Sb$_2$Te$_3$: Topologically Spin-polarized Dirac fermions with $\pi$ Berry's Phase}

\author{D. Hsieh}
\affiliation{Joseph Henry Laboratories of Physics, Princeton
University, Princeton, NJ 08544, USA}
\author{Y. Xia}
\affiliation{Joseph Henry Laboratories of Physics, Princeton
University, Princeton, NJ 08544, USA}
\author{D. Qian}
\affiliation{Joseph Henry Laboratories of Physics, Princeton
University, Princeton, NJ 08544, USA}
\author{L. Wray}
\affiliation{Joseph Henry Laboratories of Physics, Princeton
University, Princeton, NJ 08544, USA}
\author{J. H. Dil}
\affiliation{Swiss Light Source, Paul Scherrer Institute, CH-5232,
Villigen, Switzerland} \affiliation{Physik-Institut, Universit\"{a}t
Z\"{u}rich-Irchel, 8057 Z\"{u}rich, Switzerland}
\author{F. Meier}
\affiliation{Swiss Light Source, Paul Scherrer Institute, CH-5232,
Villigen, Switzerland} \affiliation{Physik-Institut, Universit\"{a}t
Z\"{u}rich-Irchel, 8057 Z\"{u}rich, Switzerland}
\author{J. Osterwalder}
\affiliation{Physik-Institut, Universit\"{a}t Z\"{u}rich-Irchel,
8057 Z\"{u}rich, Switzerland}
\author{L. Patthey}
\affiliation{Swiss Light Source, Paul Scherrer Institute, CH-5232,
Villigen, Switzerland}
\author{A. V. Fedorov}
\affiliation{Advanced Light Source, Lawrence Berkeley National
Laboratory, Berkeley, CA 94720, USA}
\author{H. Lin}
\affiliation{Department of Physics, Northeastern University, Boston,
MA 02115, USA}
\author{A. Bansil}
\affiliation{Department of Physics, Northeastern University, Boston,
MA 02115, USA}
\author{D. Grauer}
\affiliation{Department of Chemistry, Princeton University,
Princeton, NJ 08544, USA}
\author{Y. S. Hor}
\affiliation{Department of Chemistry, Princeton University,
Princeton, NJ 08544, USA}
\author{R. J. Cava}
\affiliation{Department of Chemistry, Princeton University,
Princeton, NJ 08544, USA}
\author{M. Z. Hasan}
\affiliation{Joseph Henry Laboratories of Physics, Princeton
University, Princeton, NJ 08544, USA}

\date{Submitted to PRL on 18-June, 2009}


\begin{abstract}
We show that the strongly spin-orbit coupled materials Bi$_2$Te$_3$
and Sb$_2$Te$_3$ (non-Bi Topological insulator) and their derivatives belong to the $Z_2$ (Time-Reversal-Protected) topological-insulator class. Using a combination of first-principles theoretical calculations and photoemission spectroscopy, we directly
show that Bi$_2$Te$_3$ is a large spin-orbit-induced indirect bulk
band gap ($\delta\sim150$ meV) semiconductor whose surface is
characterized by a single topological \textit{spin-Dirac cone}. The electronic
structure of self-doped Sb$_2$Te$_3$ exhibits similar $Z_2$ topological
properties. We demonstrate that the dynamics of surface spin-only Dirac fermions can be controlled through systematic Mn doping, making these materials classes
potentially suitable for exploring novel topological physics. We emphasize (theoretically and experimentally) that the Dirac node is well within the bulk-gap and not degenerate with the bulk valence band.
\end{abstract}

\maketitle

Topological insulators are a new phase of quantum matter that host
exotic Dirac electrons at their edges owing to a combination of
relativistic and quantum entanglement effects \cite{Day}. They were
recently proposed \cite{Fu:STI2,Moore:STI1,Roy} and shortly
afterwards discovered in the Bi$_{1-x}$Sb$_x$ \cite{Hsieh_Nature,
Hsieh_Science} and Bi$_2$Se$_3$ \cite{Xia, Hor} materials. In these
systems, spin-orbit coupling (SOC) gives rise to electrically
insulating states in the bulk and robust conducting states along the
edges. In contrast to graphene, which has four Dirac cones (2 doubly
degenerate cones at the K and K$'$ points in momentum space)
\cite{Novoselov(Graphene)}, the remarkable property of topological
edge states is that their dispersion is characterized by an odd
number of non-degenerate Dirac cones. Such odd spin-Dirac cone edge
metals exhibit a host of unconventional properties
including a fractional (half-integer) quantum Hall effect
\cite{Fu:STI2,Qi_QFT, zhang} and immunity to Anderson localization due to
spin-texture and $\pi$ Berry's phases on their surfaces
\cite{Fu:STI2,Hsieh_Nature,Hsieh_Science,Xia,Schnyder}. The $\pi$ Berry's phase and topological spin-textures were observed in several topological insulators and parent materials such as Bi-Sb, pure-Sb (pure Antimony), Bi$_2$Se$_3$ and Bi$_2$Te$_3$ by Xia et.al.,\cite{Xia} and Hsieh et.al.,\cite{Hsieh_Science} by spin-resolved measurements. Interesting physics may also occur at the interface between a topological insulator and an ordinary ferromagnet or superconductor, where electromagnetic responses that defy Maxwell's equations \cite{Essin,Franz,Qi_QFT} and excitations that obey non-Abelian statistics \cite{Fu_Majorana2,Akhmerov} are theoretically expected.

\begin{figure}
\includegraphics[scale=0.36,clip=true, viewport=-0.0in 0in 10.5in 8.5in]{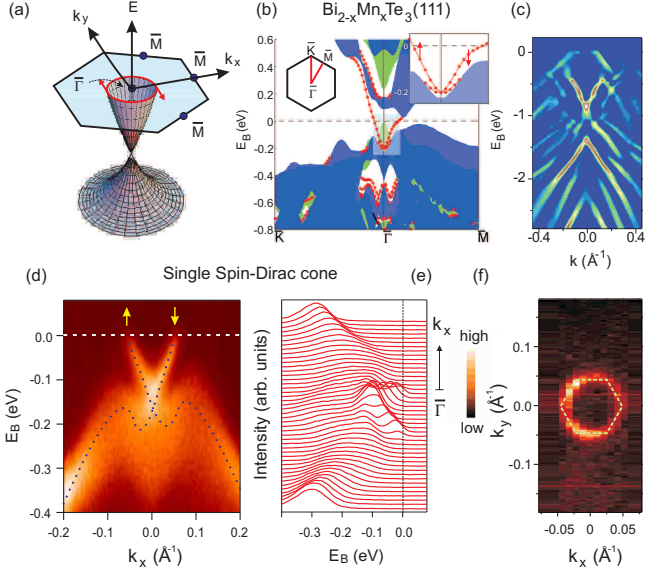}
\caption{\label{fig:STI}. \textbf{A single massless spin-Dirac cone on
the surface of Bi$_2$Te$_3$:} (a) Schematic of the (111) surface
Brillouin zone with the four time-reversal-invariant momenta
($\bar{\Gamma}$,3$\times$\={M}) marked by blue circles. A single
Fermi surface enclosing $\bar{\Gamma}$ that arises from a Dirac cone
is the signature of the most basic topological insulator. Red arrows
denote direction of spin. Spin data is taken from spin-ARPES \cite{Hsieh_Science} (b) Calculated band structure along the \={K}-$\bar{\Gamma}$-\={M} cut of the Bi$_2$Te$_3$(111) BZ. Bulk
band projections are represented by the shaded areas. The band
structure results with spin-orbit coupling (SOC) are presented in
blue and that without SOC in green. The magnitude of the bulk
indirect gap is typically underestimated by $ab$ $initio$
calculations. No pure surface band is observed within the bulk band
gap without SOC (black lines). One pure gapless surface band
crossing $E_F$ is observed when SOC is included (red lines). Inset
shows enlargement of low energy region (shaded box) near
$\bar{\Gamma}$. (c) ARPES second derivative image of the bulk
valence bands of Bi$_2$Te$_3$ along $\bar{\Gamma}$-\={M}. (d) ARPES
intensity map of the gapless surface state bands imaged one hour
after cleavage. The blue dotted lines are guides to the eye showing
the Dirac dispersion. The spin directions are marked based on
calculations. (e) Energy distribution curves of the data shown in
(d). (f) Constant energy ARPES intensity map of Bi$_2$Te$_3$
collected at $E_F$ using $h\nu$ = 35 eV. Dashed lines are guides to
the eye.}
\end{figure}

The surging number of interesting experimental proposals involving
odd Dirac cone surface metals \cite{Qi_QFT, Moore_exciton,
Fu_Majorana2, Akhmerov, Ran_dislocation} has ignited a search for
the most elementary form of a topological insulator, namely one with
a large bulk band gap and a single surface Dirac cone. Although
Bi$_{1-x}$Sb$_x$ has a room temperature direct band gap ($\delta$
$>$ 30 meV) \cite{Hsieh_Nature}, a small effective mass of its bulk
electrons is known to cause the system to form conducting impurity
bands even in high purity samples \cite{Lenoir}, which dominate over
conduction through the surface states. More importantly,
Bi$_{1-x}$Sb$_x$ has multiple surface states of both topological and
non-topological origin \cite{Hsieh_Nature}, which makes isolating
any transport signal from a single topological surface state
particularly challenging. More recently, angle-resolved
photoemission spectroscopy (ARPES) \cite{Xia} and theoretical \cite{Xia,Zhang}
evidence suggest that Bi$_2$Se$_3$ is a large band gap ($\sim$300 meV) single spin-Dirac cone topological insulator. In this Letter, we report a bulk and surface ARPES investigation of single crystals of Bi$_2$Te$_3$, Bi$_{2-x}$Mn$_{x}$Te$_3$ and Sb$_2$Te$_3$. Remarkably, we find that their electronic structures are in close agreement with our topological
SOC calculations, and a spin-Dirac cone is realized on their (111) surfaces. We emphasize (theoretically and experimentally) that the Dirac node is well within the bulk-gap and not degenerate with the bulk valence band. Although Sb$_2$Te$_3$ is found to have stable bulk states, we show that the Fermi energy of Bi$_2$Te$_3$ is time dependent, which has also been observed with ARPES in hole doped Bi$_2$Te$_3$ samples \cite{Noh}, and can be controlled via Mn doping. Using a synchrotron light source with a variable photon energy ($h\nu$), we show that the bulk-like states of Bi$_{2-x}$Mn$_{x}$Te$_3$ ($x$=0) are insulating with the valence band maximum lying around 150 meV below $E_F$, realizing a large band gap topological insulator with tuneable surface dynamics that can be used in future transport based searches for novel topological physics.

\begin{figure}
\includegraphics[scale=0.3,clip=true, viewport=-0.4in 0in 8.5in 11.0in]{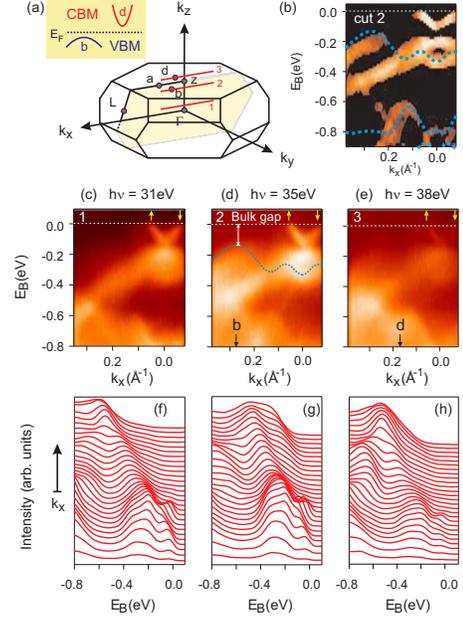}
\caption{\label{fig:Bulk}. \textbf{Observation of insulating
bulk-like states in stoichiometric Bi$_2$Te$_3$ supporting a
six-peak electronic structure:} (a) Bulk rhombohedral Brillouin zone
of Bi$_2$Te$_3$. According to LDA band structure calculations
\cite{Youn,Mishra}, six valence band maxima are located at the
\textbf{b} points that are related to one another by 60$^{\circ}$
rotations about $\hat{z}$. The red lines show the momentum space
trajectories of the ARPES scans taken using $h\nu$ = 31 eV, $h\nu$ =
35 eV and $h\nu$ = 38 eV. Inset shows a schematic of the indirect
bulk band gap. (b) Calculated valence band structure along cut 2
superimposed on the second derivative image of a corresponding ARPES
cut. The calculated band energies have been shifted downwards to
match the data. (c) to (e) show ARPES intensity maps along the
$h\nu$ = 31 eV, 35 eV and 38 eV trajectories respectively, obtained
one hour after sample cleavage. The in-plane momentum component of
the \textbf{b} and \textbf{d} points are marked by black arrows, and
the energy of the valence band maximum relative to $E_F$ ($\delta$)
is shown by the double-headed arrow. Yellow arrows denote direction
of spin of Dirac cone surface states. (f) to (h) show the energy
distribution curves corresponding to images (c) to (e)
respectively.}
\end{figure}

ARPES measurements were performed with 28 to 45 eV linearly
polarized photons on beam line 12.0.1 at the Advanced Light Source
in Lawrence Berkeley National Laboratory. The typical energy and
momentum resolution was 15 meV and 1\% of the surface Brillouin zone
(BZ) respectively. Single crystals of Bi$_{2-x}$Mn$_x$Te$_3$ were
grown by melting stoichiometric mixtures of elemental Bi (99.999
\%), Te (99.999 \%) and Mn (99.95 \%) at 800$^{\circ}$C overnight in
a sealed vacuum quartz tube. The crystalline sample was cooled over
a period of two days to 550$^{\circ}$C, and maintained at the
temperature for 5 days. The same procedure was carried out with Sb
(99.999 \%) and Te (99.999 \%) for Sb$_2$Te$_3$ crystals. Samples
were cleaved in ultra high vacuum (UHV) at pressures better than
$5\times10^{-11}$ torr at 15 K. Our calculations were performed with
the linear augmented-plane-wave method in slab geometry using the
WIEN2K package \cite{Blaha}. Generalized gradient approximations of
Perdew, Burke, and Ernzerhof \cite{Perdew} was used to describe the
exchange-correlation potential. Spin-orbit coupling was included as
a second variational step using scalar-relativistic eigenfunctions
as basis. The surface was simulated by placing a slab of six
quintuple layers in vacuum using optimized lattice parameters from
\cite{Wang}. A grid of 35$\times$35$\times$1 points were used in the
calculations, equivalent to 120 k-points in the irreducible BZ and
2450 k-points in the first BZ.

\begin{figure*}
\includegraphics[scale=0.35,clip=true, viewport=-0.8in 0in 12.5in 4.6in]{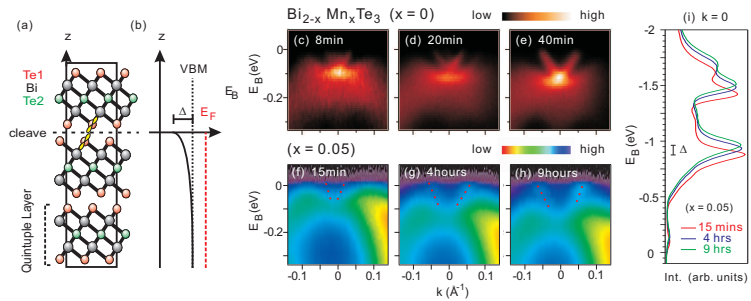}
\caption{\label{fig:Time}. \textbf{Slow dynamics of the surface
Dirac cone dispersion in Bi$_{2-x}$Mn$_x$Te$_3$:} (a) The crystal
structure of Bi$_2$Te$_3$ viewed parallel to the quintuple layers.
The Te(1) 5$p$ orbitals that form the inter-quintuple layer van der
Waals bonds are shown in yellow. (b) Schematic of the band bending
of the bulk valence band maximum (VBM) near the cleaved surface. (c)
ARPES spectra of Bi$_2$Te$_3$ along the $\bar{\Gamma}$-\={M}
direction taken with $h\nu$ = 30 eV (c) 8 mins, (d) 20 mins, and (e)
40 mins after cleavage in UHV. Analogous ARPES spectra for
Bi$_{1.95}$Mn$_{0.05}$Te$_3$ (f) 15 mins, (g) 4 hours and (h) 9
hours after cleavage, showing a slower relaxation rate. Red lines
are guides to the eye. (i) The energy distribution curves of
Bi$_{1.95}$Mn$_{0.05}$Te$_3$ at $\bar{\Gamma}$ taken at different
times after cleavage.}
\end{figure*}

The most basic 3D topological insulator supports a single Dirac cone
on its surface (Fig.~\ref{fig:STI}(a)), with the Dirac node located
at a momentum \textbf{k}$_T$ in the surface Brillouin zone (BZ),
where \textbf{k}$_T$ satisfies \textbf{k}$_T$ = -\textbf{k}$_T$ +
\textbf{G} and \textbf{G} is a surface reciprocal lattice vector
\cite{Fu:STI2}. Our theoretical calculations on Bi$_2$Te$_3$ (111)
show that it is a SOC induced bulk band insulator, and that a single
surface Dirac cone that encloses \textbf{k}$_T$ = $\bar{\Gamma}$
only appears when SOC is included (Fig.~\ref{fig:STI}(b)). To
determine whether single crystalline Bi$_2$Te$_3$ is a topological
insulator as predicted, we first mapped its high energy valence
bands using ARPES. Figure~\ref{fig:STI}(c) and Figure 2 show that
the measured bulk band structure is well described by SOC
calculations, suggesting that the electronic structure is
topologically non-trivial. A more direct probe of the topological
properties of Bi$_2$Te$_3$, however, is to image its surface states.
Figure~\ref{fig:STI}(d) and (e) show that the surface states are
metallic and are characterized by a single Dirac cone crossing
E$_F$, in agreement with theory (Fig.~\ref{fig:STI}(b)). Moreover,
the density of states at E$_F$ is distributed about a single ring
enclosing $\bar{\Gamma}$ (Fig.~\ref{fig:STI}(f)), in accordance with
Bi$_2$Te$_3$ being a topological insulator.

\begin{figure}
\includegraphics[scale=0.7,clip=true, viewport=-0.0in 0in 5.0in 7.0in]{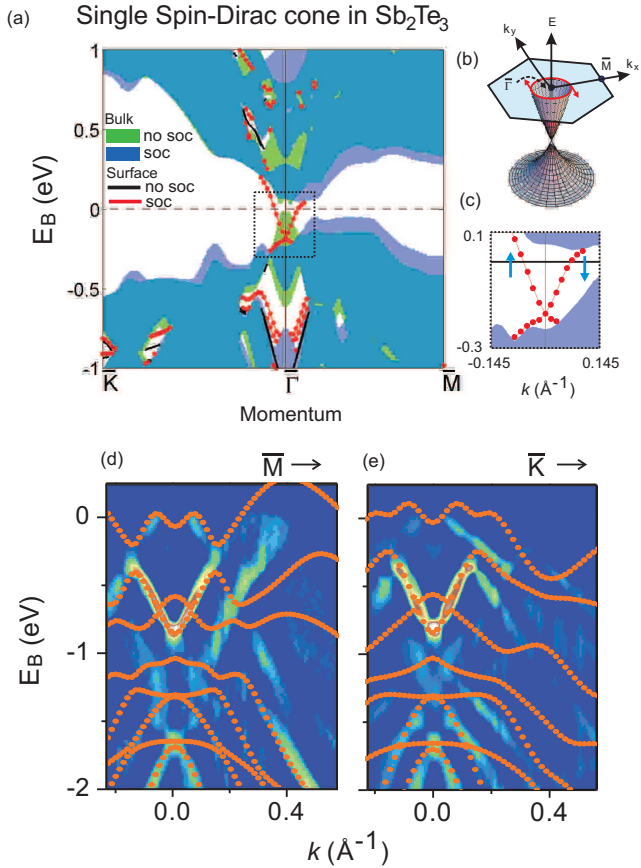}
\caption{\label{fig:SbTe}. \textbf{Evidence for a topologically
non-trivial band structure and spin-Dirac cone in Sb$_2$Te$_3$ :} (a) Calculated band
structure along the \={K}-$\bar{\Gamma}$-\={M} cut of the
Sb$_2$Te$_3$(111) BZ. Bulk band projections are represented by the
shaded areas. The bulk (surface) band structure results with
spin-orbit coupling (SOC) are presented in blue (red lines) and that
without SOC in green (black lines). (b) Schematic of the single
surface Dirac cone in Sb$_2$Te$_3$ based on calculations. (c)
Enlargement of low energy region (shaded box in (a)) near
$\bar{\Gamma}$. (d) Second derivative image of the bulk valence
bands along $\bar{\Gamma}$-\={M} and (e) $\bar{\Gamma}$-\={K} at
$k_z$ = -0.77 $\Gamma-Z$. The superposed SOC calculation (orange
dots) has been rigidly shifted upwards by 300 meV to match the
data.}
\end{figure}

Our theoretical calculations show that stoichiometric Bi$_2$Te$_3$
is a bulk indirect gap insulator (Fig.~\ref{fig:STI}(b)). The bulk
valence band maximum (VBM) in Bi$_2$Te$_3$ lies at the
\textbf{b}-point in the $\Gamma Z L$ plane of the three-dimensional
bulk BZ (Fig.\ref{fig:Bulk}(b)), giving rise to VBM in each of six
such mirror planes in agreement with previous proposals
\cite{Youn,Mishra}. The VBM exhibits an indirect gap with the
conduction band minimum (CBM) above E$_F$, which is located at the
\textbf{d}-point in the $\Gamma Z L$ plane. In order to establish
whether Bi$_2$Te$_3$ is a bulk insulator as predicted, we performed
a series of ARPES scans along the cuts shown by red lines in
Figure~\ref{fig:Bulk}(a) (displaced along $k_z$ by varying the
incident photon energy) that traverse the locations of the VBM and
CBM in the bulk BZ. All $h\nu$-dependent scans were taken more than
an hour after cleavage to allow the band structure to stabilize (see
Fig.~\ref{fig:Time}). Figures~\ref{fig:Bulk}(c)-(h) show a series of
ARPES band dispersions along momentum cuts in the $k_x$-$k_z$ plane
taken using photon energies of 31 eV, 35 eV and 38 eV respectively.
The Dirac cone near E$_F$ shows no dispersion with $h\nu$,
supporting its surface state origin. In contrast, a strongly $h\nu$
dispersive hole-like band is observed near $k_x$ = 0.27 \AA$^{-1}$,
whose maximum rises to an energy $\delta$ closest to E$_F$ ($\delta$
= $-150\pm$50 meV) when $h\nu$ = 35 eV (Fig.~\ref{fig:Bulk}(d)).
Using the free electron final state approximation, the VBM is
located at (0.27, 0, 0.27) \AA$^{-1}$, in agreement with
calculations. ARPES scans taken in the vicinity of the
\textbf{d}-point (0.17, 0, 0.37) \AA$^{-1}$, which is traversed
directly when $h\nu$ = 38 eV, do not measure any signal from the
CBM, showing that E$_F$ lies in the bulk band gap. This is
consistent with the size of the indirect band gap ($>$ 150 meV)
measured using tunneling \cite{Urazhdin} and optical techniques
\cite{Thomas}. We note that because ARPES is only sensitive to the
topmost quintuple layer (Fig.~\ref{fig:Time}(a)) at our sampled
photon energies \cite{Hufner}, the measured energy of the bulk band
edge $\delta$ may differ from the true bulk value due to band
bending effects that are commonly observed in semiconductors.

In order to investigate the effects of semiconductor band bending on
the surface Dirac cone on Bi$_2$Te$_3$, we performed time dependent
ARPES experiments. Our results show that the binding energy of the
Bi$_2$Te$_3$ surface Dirac node exhibits a pronounced time
dependence, increasing from $E_B \sim$ -100 meV 8 minutes after
cleavage to $E_B \sim$ -130 meV at 40 minutes
(Fig.~\ref{fig:Time}(c)-(e)), in agreement with a previous report
\cite{Noh}. Such behavior has been attributed to a downward band
bending near the surface (Fig.~\ref{fig:Time}(b)) that is caused by
the breaking of inter-quintuple layer van der Waals Te(1)-Te(1)
bonds (Fig.~\ref{fig:Time}(a)), which creates a net electric field
near the surface upon crystal termination \cite{Urazhdin, Mishra}.
Unlike previous calculations \cite{Zhang}, our calculated position
of the Dirac node lies in the bulk band gap (Fig.~\ref{fig:STI}(b)),
which corroborates our experimental finding that the intensity is
strongest near the Dirac node and drastically weakens away from
$\bar{\Gamma}$ as the surface band merges with the bulk bands and
become short-lived \cite{Hufner,Hsieh_Nature,Hsieh_Science}. The
slow dynamics of the band bending process suggests that charge
accumulation at the surface is coupled to a much slower surface
lattice relaxation \cite{Noh}. The system is likely to be
significantly delayed in achieving equilibrium by local
lattice/charge density fluctuations such as may arise from site
defects, which are prominent in such materials \cite{Urazhdin,Hor}.
By systematically increasing the defect concentration through Mn for
Bi substitution, we demonstrate here that band bending can be slowed
by up to 10 fold (Fig.~\ref{fig:Time}(f)-(h)), allowing a wider
range of the intrinsic relaxation time scale to be accessed. ARPES
valence band spectra (Fig.~\ref{fig:Time}(i)) of
Bi$_{1.95}$Mn$_{0.05}$Te$_3$ taken over a 15 hour period show that
the positions of the valence band edges shift downward by a total
energy of around 100 meV, which we take as a measure of the total
magnitude of band bending $\Delta$.

Having identified the topological insulator Bi$_2$Te$_3$, we
proceed to investigate whether similar topological effects can take
place in a non bismuth based compound. Figure~\ref{fig:SbTe}(a)
shows the calculated electronic structure of Sb$_2$Te$_3$, which,
like Bi$_2$Te$_3$, exhibits a bulk insulating band structure that is
strongly influenced by SOC and a single Dirac cone on its (111)
surface. By comparing our SOC calculations with the experimentally
measured bulk valence bands, it is clear that there is good
agreement along both the $k_x$ (Fig.~\ref{fig:SbTe}(d)) and $k_y$
(Fig.~\ref{fig:SbTe}(e)) directions, showing that the bulk
electronic structure of Sb$_2$Te$_3$ is consistent with having
topologically non-trivial bulk properties. However, due to a high
level of intrinsic doping that is typical of these compounds
\cite{Hor,Xia}, the Fermi energy of naturally grown Sb$_2$Te$_3$
lies in the bulk valence band continuum and thus does not cut
through the surface states. Unlike Bi$_{2-x}$Mn$_{x}$Te$_3$, no time
dependence of the bands is observed. Recently, we came across independent work on the Bi$_2$(Sn)Te$_3$ (Sn-doping\cite{Chen}) series that finds a single Dirac cone on the surface. Single spin-Dirac cone and Berry's phase on these classes of materials were first presented in \cite{Xia} [also see \cite{Note Added}].

In conclusion, our first-principles theoretical predictions and calculations and
photoemission results show that Bi$_2$Te$_3$ and Sb$_2$Te$_3$ possess bulk
band structures where the insulating gap originates from a large
spin-orbit coupling term, and such insulators support topologically
nontrivial $Z_2$ (Time-Reversal-Protected nature and the absence of backscattering are guaranteed by the odd number of spin-polarized crossings as the character of $Z_2$ topology) surface states. Our direct observation of single Dirac cones in these materials and the systematic methods demonstrated to control the Dirac fermion dynamics on these highly non-trivial surfaces point to new opportunities for spintronic and
quantum-information materials research.

Note Added: The experimental data pre-existed the theoretical calculations. The surface-state data and spin-ARPES methods were presented at two KITP conference proceedings (BiSb, Bi$_2$Te$_3$, Sb$_2$Te$_3$, pure Sb and Bi$_2$Se$_3$) were presented in two talks : Direct Determination of Topological Order:Topological Quantum Numbers and Berry's Phase from Spin-Texture Maps of Spin-Orbit Insulators. See $http://online.itp.ucsb.edu/online/motterials07/hasan/$ (2007) and $http://online.itp.ucsb.edu/online/qspinhall\_m08/hasan/$ (2008).


\begin{thebibliography}{99}




\bibitem{Day}
C. Day. \textit{Physics Today} \textbf{62}, 12 (2009).


\bibitem{Fu:STI2}
L. Fu, C. L. Kane and E. J. Mele. \textit{Phys. Rev. Lett.}
\textbf{98}, 106803 (2007).

\bibitem{Moore:STI1}
J. E. Moore and L. Balents. \textit{Phys. Rev. B} \textbf{75}
121306(R) (2007).

\bibitem{Roy}
R. Roy. arXiv:cond-mat0604211v2 (2006).

\bibitem{zhang} S.-C. Zhang, Physics \textbf{1}, 6 (2008); C.L. Kane, Nature Phys. \textbf{4}, 348 (2008).

\bibitem{Hsieh_Nature}
D. Hsieh $et$ $al.$ \textit{Nature} (London) \textbf{452}, 970 (2008).

\bibitem{Hsieh_Science}
D. Hsieh $et$ $al.$, \textit{Science} \textbf{323}, 919 (2009); D. Hsieh $et$ $al.$, \textit{Nature} (London) 460, 1101 (2009).
http://dx.doi.org/10.1038/nature08234 (2009).

\bibitem{Xia}
Y. Xia $et$ $al$. \textit{Nature Phys.} \textbf{5}, 398 (2009).

\bibitem{Hor}
Y. S. Hor $et$ $al.$ \textit{Phys. Rev. B} \textbf{79}, 195208
(2009).

\bibitem{Novoselov(Graphene)}
K. S. Novoselov $et$ $al.$ \textit{Nature} (London) \textbf{438}, 197 (2005).

\bibitem{Qi_QFT}
X.-L. Qi, T. L. Hughes and S.-C. Zhang. \textit{Phys. Rev. B}
\textbf{78}, 195424 (2008).

\bibitem{Schnyder}
A. P. Schnyder, S. Ryu, A. Furusaki and A. W. W. Ludwig.
\textit{Phys. Rev. B} \textbf{78} 195125 (2008).


\bibitem{Essin}
A. Essin, J. E. Moore and D. Vanderbilt. \textit{Phys. Rev. Lett.}
\textbf{102}, 146805 (2009).

\bibitem{Franz}
M. Franz. \textit{Physics} \textbf{1}, 36 (2008).

\bibitem{Fu_Majorana2}
L. Fu and C. L. Kane. \textit{Phys. Rev. Lett.} \textbf{102}, 216403
(2009).

\bibitem{Akhmerov}
A. R. Akhmerov, J. Nilsson and C. W. J. Beenakker. \textit{Phys.
Rev. Lett.} \textbf{102}, 216404 (2009).

\bibitem{Ran_dislocation}
Y. Ran, Y. Zhang and A. Vishwanath. \textit{Nature Phys.}
\textbf{5}, 298 (2009).

\bibitem{Moore_exciton}
B. Seradjeh, J. E. Moore and M. Franz. \textit{Phys. Rev. Lett.}
\textbf{103}, 066402 (2009).

\bibitem{Lenoir}
B. Lenoir, M. Cassart, J.-P. Michenaud, H. Scherrer and S. Scherrer.
\textit{J. Phys. Chem. Solids}. \textbf{57}, 89 (1996).

\bibitem{Zhang}
H. Zhang $et$ $al$. \textit{Nature Phys.} \textbf{5}, 438 (2009).

\bibitem{Noh} First observation of linear Dirac bands in Bi$_2$Te$_3$ were reported by 
H.-J. Noh $et$ $al.$ \textit{Europhys. Lett.} \textbf{81}, 57006
(2008). No spin-polarization or pi Berry's phase, critical for proving topological order were observed here.

\bibitem{Blaha}
P. Blaha $et$ $al.$ Computer code WIEN2K. Vienna University of
Technology, Vienna (2001).

\bibitem{Perdew}
J.P. Perdew, K. Burke, \& M. Ernzerhof \textit{Phys. Rev. Lett.}
\textbf{77}, 3865-3868 (1996).

\bibitem{Wang}
G. Wang, \& T. Cagin \textit{Phys. Rev. B} \textbf{76}, 075201
(2007).

\bibitem{Youn}
S. J. Youn and A. J. Freeman. \textit{Phys. Rev. B} \textbf{63},
085112 (2001).

\bibitem{Mishra}
S. K. Mishra, S. Satpathy and O. Jepsen. \textit{J. Phys: Condens.
Mat.} \textbf{9}, 461 (1997).


\bibitem{Urazhdin}
S. Urazhdin $et$ $al.$ \textit{Phys. Rev. B} \textbf{69}, 085313
(2004).

\bibitem{Thomas}
G. A. Thomas $et$ $al$. \textit{Phys. Rev. B} \textbf{46}, 1553
(1992).

\bibitem{Hufner}
S. H{\"{u}}fner. Photoelectron Spectroscopy. Springer, Berlin
(1995).



\bibitem{Chen}
Y.L. Chen $et$ $al$. arXiv:0904.1829v1 [ScienceExpress, 11-June, 2009].

\bibitem{Note Added} The surface-state data (BiSb, Bi$_2$Te$_3$, pure Sb, Sb$_2$Te$_3$ and Bi$_2$Se$_3$) and the unique spin-ARPES methods were presented at the KITP conference proceedings in (2007, 2008): $http://online.itp.ucsb.edu/online/motterials07/hasan/$ (2007) and $http://online.itp.ucsb.edu/online/qspinhall\_m08/hasan/$
    (Direct Determination of Topological Order:Topological Quantum Numbers and Berry's Phase from Spin-Texture Maps of Spin-Orbit Insulators)


\end{thebibliography}

\end{document}